\documentclass[aps,pre,twocolumn,superscriptaddress,showpacs]{revtex4}
\usepackage{amssymb}
\usepackage{epsfig}
\usepackage{amsmath}
\usepackage{times}
\usepackage{graphicx}
\usepackage{amscd}
\usepackage{amsfonts}
\usepackage{fancyhdr}
\usepackage{subfigure}

\begin{document}

\title{Opinion Dynamic with agents immigration}

\author{Zhong-Lin Han}
\affiliation{Department of Physics, University of Science and
Technology of China, Hefei 230026, China}

\author{Yu-Jian Li}
\affiliation{Department of Modern Physics, University of Science and
Technology of China, Hefei 230026, China}

\author{Bing-Hong Wang}
\affiliation{Department of Modern Physics, University of
Science and Technology of China, Hefei 230026, China}

\begin{abstract}
Abstract
\end{abstract}

\date{\today}


\maketitle

\section{Introduction} \label{sec:intro}
Recent years, a large class of interdisciplinary problems has been
successfully studied with statistical physics methods. Statistical
physics establishes the bridge from microscopic characteristics to
macroscopic behaviors, for systems containing a large number of
interacting components. Using both analytical and numerical tools,
it has contributed greatly to our understanding of various complex
systems. In this paper, we are motivated by the statistical physics
of a sociological problem, namely, opinion dynamics.

As one of the classical and traditional research areas in both
social science and theory physics, opinion dynamics has attracted
much attention. A lot of models concerning the process of opinion
formation, such as voter model, bounded confidence model, have been
proposed previously. Meanwhile, some of recent studies discussed and
described the opinion dynamics on both common conditions and various
complex networks.

The issue of individual mobility has become increasingly fundamental
due to the Human migration and human dynamic. The issue is also
important in other contexts such as the emergence of Cooperation
among individuals [20] and species coexistence in cyclic competing
games [21]. Recently, some empirical data of human movements have
been collected and analyzed [22,23]. From the standpoint for dynamic
of complex systems, when individuals (nodes, agents) are mobile, the
edges in the topological structures are no longer fixed, yielding
more different results on that than before.

In our paper, we try to propose a new model combining conventional
opinion dynamics with agents immigration according to information
transmission and evolution. In our simulation, we finally find a
series of results reflecting special and different features of
opinion dynamics with immigration. By introducing a parameter
$\alpha$ to control the weight of influence of individual opinions,
according to a recent study considering weight influence, we find
that there also exist an optimal value of $\alpha$ leading to the
shortest consensus time for all individuals on a isotropic plane we
concern. After presenting the results of simulation in different
situations, we also analysis the results of our model in
mathematical way, which leads us finding out what are the exactly
direct factors impacting the exponent of weight of individual
opinions.

\begin{figure}
\begin{center}
\epsfig{figure=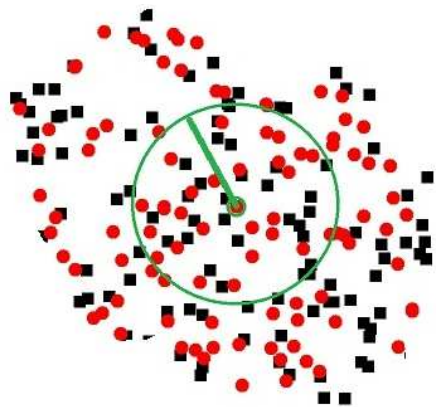,width=6cm} \caption{(Color online) For $N =
2000$ and $\langle k \rangle=4$, the density $\rho_{c}$ as a
function of $\alpha$ for different values of $r$ in the case where
all cooperators contribute the same cost c per game. Every
cooperator contributes a cost $c=1$ in every neighborhood that it
plays.} \label{fig:seclimit}
\end{center}
\end{figure}

In this paper we found up a new type of model for information
dynamics with immigration and we state related parameters and rules
of our model. In order to demonstrate the rationality of it, we
presented both computational simulation and mathematical analysis.
Another most significant thing is that we have designed a new
mathematical method for model with linear algebra.Compared with the
previous method,we finally finish a complete model for opinion and
information dynamics with immigration.

\section{Model} \label{sec:theory}
As previous classical model focusing on the material process of
spread and formation of opinions, we spend more efforts on finding
special results when individuals carry opinions with immigration. To
focus on a more efficient situation, we just confine our discussion
on an isotropic plane, without special network effects. On the other
hand, the individuals we concern are just holding two kinds of
opinion, the positive opinion $\psi_+>0$ and the negative opinion
$\psi_- <0$. According to one model on opinion dynamics proposed
before (.), we introduce the weight exponent to control the weight
of influence of each individual. We describe that all of the
opinions of individuals evolve simultaneously completely rely on its
neighbors' opinion and neighbors' weight. Here we describe the
evolution process of the whole individuals on the plane in
mathematical way, which could be denoted as follow (yang han xing)

\begin{eqnarray}
p_{+}= \frac{\sum_i^u \omega_i^\alpha}{\sum_i^u \omega_i^\alpha +
\sum_j^v \omega_j^\alpha },
\end{eqnarray}
\begin{eqnarray}
p_{-}=\frac{\sum_i^v \omega_i^\alpha}{\sum_i^u \omega_i^\alpha +
\sum_j^v \omega_j^\alpha }
\end{eqnarray}

where $p_{+}$ and $p_{-}$ denote the probability of choosing
positive opinion and negative opinion,and the number of  neighbors
holding positive opinion is $u$ while the number of negative ones is
$v$. Here the model considers the agents with weight impact
$\omega_{i}$, which is controlled by weight exponent $\alpha$. If
the probability $p_{+}$ is lager than $p_{-}$, the agent we concern
will choose positive opinion at the next step. And it will be same
as choosing negative opinion.

\begin{figure}
\begin{center}
\epsfig{figure=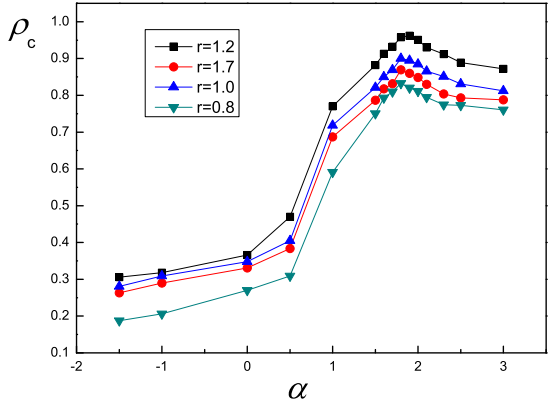,width=\linewidth} \caption{(Color online)
Cumulative payoff distribution for different values of $\alpha$. The
distribution is obtained after the cooperation density becomes
stable. The multiplication factor is set to be $r=1.6$. Solid curves
are theoretical predictions from Eq.~(\ref{eq:wealthdis}). }
\label{fig:wealthdis}
\end{center}
\end{figure}

In this model, the individual we concern evolves its opinion at
$t+1$ according to its neighbors' opinion in its view radius $r$,
which is shown in Figure.1. In Figure.1, the red agents hold
positive opinions and black agents hold negative opinions. There are
$u+v$ neighbors in the view range of individual we concern, while
here are $u$ individuals hold positive opinion and $v$ individuals
hold negative opinion at $t$ step. After comparing the weight of
positive opinion and negative opinion, the individual we concern
evolves its opinion at $t+1$ step as this

\begin{equation}
\psi_i^{(t+1)}=\sum_{r}\psi_j^{(t)}
\end{equation}

where $\psi_j^{(t)}$ denotes the opinion state of the $j$th neighbor
of the $i$th agent we concern at the $t$ step.
\begin{figure}
\begin{center}
\epsfig{figure=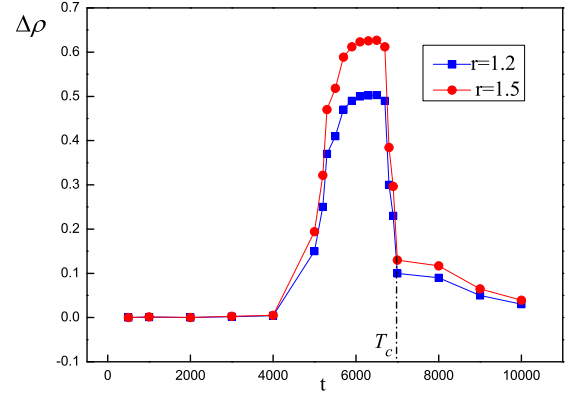,width=\linewidth} \caption{(Color online)
Times series of cooperator density in hubs' neighborhoods for (a)
The multiplication factor is $r=1.2$ and each data point is obtained
by averaging over 50 runs.} \label{fig:analysis}
\end{center}
\end{figure}
After changing their opinion in the way above, all the individuals
immigrate on the plane. All the agents would be confined in the
plane by periodic boundary condition. The velocity and direction
angle of each agent are randomly distributed, which are kept by each
agent all the time. After enough period of time, number of the
individuals holding positive opinions $N_{+}$ and the number of
other individuals holding negative ones $N_{-}$ reach a plateaus and
dynamic equilibrium. At that certain point, we believe that the
process of opinion dynamics would be terminated. And we could find
that if all of the individuals enter the plateaus, the total number
of individuals who hold positive opinion at $t$ step $\psi_+^{t}$
would be approximately equal to the number of individuals who also
hold that at $t+1$ step $\psi_+^{t+1}$. And we could carry on this
description with mathematical language,
\begin{equation}
 \sum_i^N\psi_i^{(t+1)}=\sum_i^N \psi_i^{(t)}
\end{equation}
The total time steps the system took could be defined as $T_{c}$ for
convergent time.

\subsection{Results and analysis}
In the following discussion and simulation, we confines our
individuals on an isotropic plane $(L\times L)$. The length of the
boundary of this plane $L$ is 20, and the total number of
individuals on the plane would be $N$. Here we simulate the
individuals have their initial velocity under Gauss distribution,
which would be more rational and close to facts. Each individual
hold their opinions (positive one or negative one) and their fixed
weight of opinion with random probability. The distribution of
agents' weight was established in a random way at the beginning of
evolution. The exponent $\alpha$ in equations (1) and (2) controls
the evolution process. And here we define $\rho$ and $\Delta\rho$ to
describe the changing process of the individuals holding positive
opinion. They are denoted in equations as follow
\begin{equation}
\rho_c=\frac{N_{+}}{N}
\end{equation}
\begin{equation}
\Delta \rho = \frac{\Delta N_{+}}{N}
\end{equation}
In Figure 2,we show $\rho_{c}$ as function of evolution time $t$ for
different view radius $r$, both $r$=1.2 and $r$=1.5. The most
interesting thing we could find in this figure is that when
evolution time $t$ is around 6500, the value of $\Delta\rho$
plummets obviously, which finally reach the level under 0.1. In
fact, when changes of $\Delta\rho$ has lower amplitude of variation,
it also means that the individuals holding positive opinion enter
the period of dynamic equilibrium.
\begin{figure}
\begin{center}
\epsfig{figure=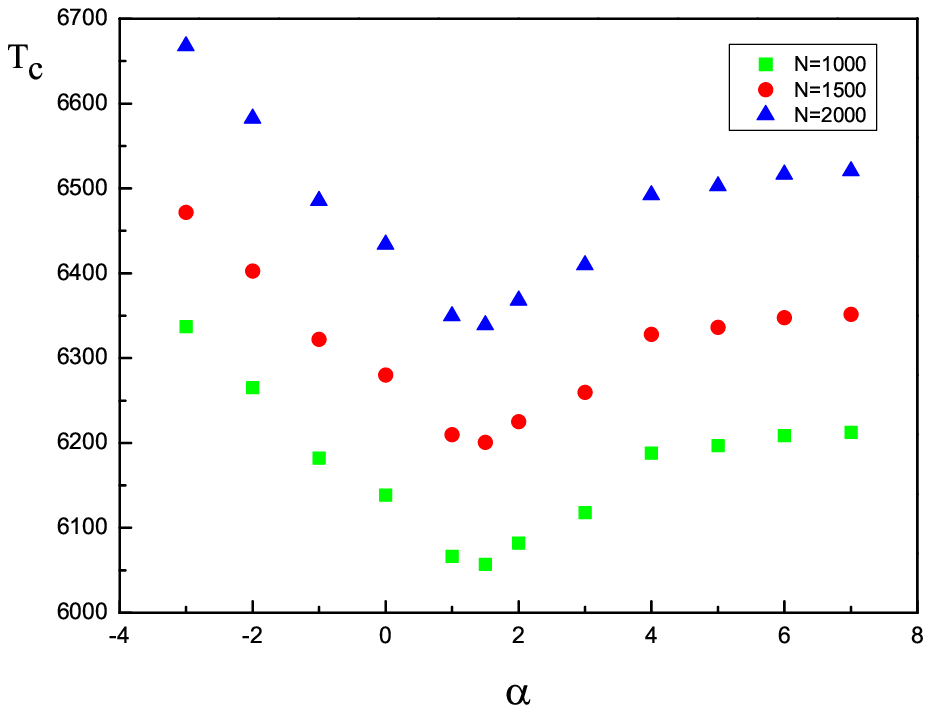,width=\linewidth} \caption{(Color online) For
$r = 1.6$, cooperator density $\rho_{c}$ as a function of degree for
different values of $\alpha$.} \label{fiq:rho_degree}
\end{center}
\end{figure}

\begin{figure}
\begin{center}
\epsfig{figure=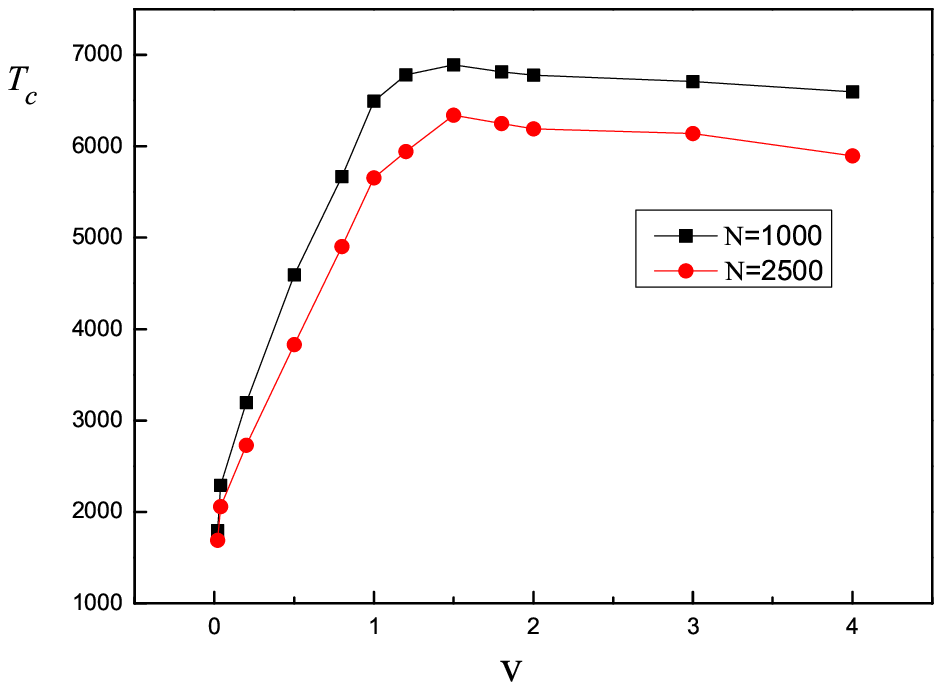,width=\linewidth} \caption{(Color online)
For $r = 1.6$, cooperator density $\rho_{c}$ as a function of degree
for different values of $\alpha$.} \label{fiq:rho_degree}
\end{center}
\end{figure}

\begin{figure}
\begin{center}
\epsfig{figure=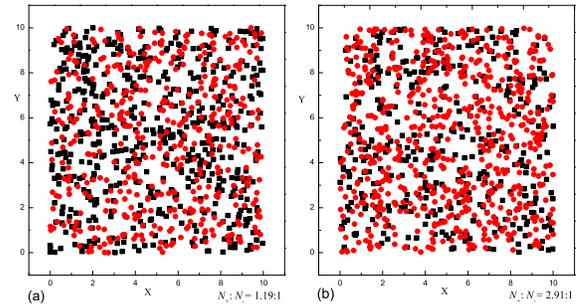,width=\linewidth} \caption{(Color online)
(a)[Initial distribution of agents with two kinds of
opinions,$N_{+}$:$N_{-}$=1.19:1. (b)Final distribution of agents
with two kinds of opinions, $N_{+}$:$N_{-}$=2.91:1.}
\label{fiq:rho_degree}
\end{center}
\end{figure}
In this situation, we could find the consensus time in Figure 3. If
$t<4000$ or $t>6500$, the $\Delta\rho$ changes in a very small range
which would also be shown in the figure. But the consensus time is
not directly impacted by only view radius $r$ of each individual. In
Figure 4, here converge time $T_{c}$ is taken as a function of
average velocity $\alpha$ , while the view radius $r$ is equal to
1.2. Interestingly, we show that would reach a minimum value when
$\alpha$ is around 2 under different values of total number of
individuals on the plane $N$. Here we present that consensus time
$T_{c}$ changes with $\alpha$ in a "smile curve". And certainly the
value of would be higher if the $N$ is more. In fact, it is obvious
to explain that when there are more individuals holding different
opinions, they would take more time to reach consensus or dynamic
equilibrium. In that we show that $N$ and weight exponent $\alpha$
could both directly determine the consensus time $T_{c}$. The more
cogent demonstration is shown in Figure 2, which presents as a
function of $\alpha$. Here $\rho_{c}$ is the density of individuals
holding positive opinions at the consensus time.

In the Figure 2, we present that $\rho_{c}$ will reach a maximum
when is around 2. Meanwhile, the value of   is greater if view
radius is lager. To show the result in a more intuitive way, we
present the distribution map in Figure 6. In Figure 6, the red
points are the ones holding positive opinions while the black points
present negative ones. In this figure, we present the specific
distribution.

To support former results, we mainly focus on the results that
positive opinions take dominant rate. In order to discuss the
parameters reflecting immigration of individuals, we present
consensus time $T_{c}$ as a function of average velocity of
individuals with different $N$ in Figure 5. In this figure, we find
that consensus time $T_{c}$ would increase approximately in a linear
way when $v$ is less than 1.5. After that  , it decreases in a
certain range without sharp changes.

To discuss the model in a more reliable way, we try to analysis the
process by founding up a series of equations for $m$ agents in total
as follow. In equations, we define that the $i$th individual we
concern has a view radius $r$, and at the t step there are $s_{m}$
individuals in its view range as its neighbors,and here we define
that $s_{ij}$ as the $i$th neighbor of the $j$th agent we concern.
In that, it is $j$th opinion state of agent's neighbor at $t$ step
that determine the opinion updating of this individuals at next time
step $t+1$. If $\psi$ is positive, individuals who holding positive
opinions would have greater weight than those who hold negative
opinions. As a consequence, the equations could be formed as follow:
\begin{widetext}
\begin{eqnarray}
 \left\{
\begin{array}{c}
\psi_{1}^{(t+1)}=\psi^{(t)}(r,s_{11})\cdot \omega_{s_{11}}^\alpha +
\ldots +\psi^{(t)}(r,s_{m-1,1})\cdot \omega_{s_{m-1,1}}^\alpha
+\psi^{(t)}(r,s_{m,1})\cdot \omega_{s_{m,1}}^\alpha \\
\vdots \\
\psi_{m-1}^{(t+1)}=\psi^{(t)}(r,s_{1,m-1})\cdot
\omega_{s_{11}}^\alpha + \ldots +\psi^{(t)}(r,s_{m-1,m-1})\cdot
\omega_{s_{m-1,m-1}}^\alpha
+\psi^{(t)}(r,s_{m,m-1})\cdot \omega_{s_{m,m-1}}^\alpha \\
\psi_{1}^{(t+1)}=\psi^{(t)}(r,s_{1,m})\cdot \omega_{s_{1,m}}^\alpha
+ \ldots +\psi^{(t)}(r,s_{m-1,m})\cdot \omega_{s_{m-1,m}}^\alpha
+\psi^{(t)}(r,s_{m,m})\cdot \omega_{s_{m,m}}^\alpha \\
\end{array}
\right.
\end{eqnarray}
\end{widetext}
To describe the model in a simpler way, we try to apply linear
algebra instead of these traditional equations. In order to write in
that way, we also introduce a new parameter $n_{ij}^{(t)}$ into this
matrix description. Here $n_{ij}^{(t)}$ reflects that the times of
opinion exchanging or sharing of $j$th individual we concern at $t$
step. In other word, $n_{ij}^{(t)}$ is a standard that concerns how
many times the $i$th individual impacts others opinion updating
choice of next time step at $t$ step. When the whole individuals get
into the plateaus of dynamic equilibrium, we discussed in Sec.2, the
opinions individuals holding would be described as where $\omega(i)$
is the weight of the ith agent, which is
\begin{widetext}
\begin{equation}
\left(
\begin{array}{c}
\psi_1^{(t+1)} \\
\psi_2^{(t+1)} \\
\vdots \\
\psi_{m-1}^{(t+1)}\\
\psi_{m}^{(t+1)}\\
\end{array}
\right)=
\left( \begin{array}{ccccc}
\psi_1^{(t)} & \psi_2^{(t)} & \ldots & \psi_{m-1}^{(t)} & \psi_m^{(t)} \\
\psi_1^{(t)} & \vdots & \vdots & \vdots & \vdots \\
\vdots & \vdots & \vdots & \vdots & \vdots \\
\vdots & \vdots & \vdots & \vdots & \psi_m^{(t)} \\
\psi_1^{(t)} & \psi_2^{(t)} & \ldots & \ldots & \psi_m^{(t)}
\end{array}
\right) \times \left( \begin{array}{cccc}
n_{11}^{(t)}\cdot \omega_{1}^\alpha & n_{12}^{(t)}\cdot \omega_{1}^\alpha & \ldots & n_{1m}^{(t)}\cdot \omega_{1}^\alpha \\
n_{21}^{(t)}\cdot \omega_{2}^\alpha & n_{22}^{(t)}\cdot \omega_{2}^\alpha & \ldots & n_{2m}^{(t)}\cdot \omega_{2}^\alpha \\
\vdots & \vdots & \ldots &\vdots  \\
n_{m-1,1}^{(t)}\cdot \omega_{m-1}^\alpha & n_{m-1,2}^{(t)}\cdot \omega_{m-1}^\alpha & \ldots & n_{m-1,m}^{(t)}\cdot \omega_{m-1}^\alpha \\
n_{m,1}^{(t)}\cdot \omega_{m}^\alpha & n_{m,2}^{(t)}\cdot \omega_{m}^\alpha & \ldots & n_{m,m}^{(t)}\cdot \omega_{m}^\alpha
\end{array}
\right)
\end{equation}
\end{widetext}
Here we denote that $n_{ij}^{(t)}$ as the times for opinion
exchanges between the $i$th and $j$th agents at the $t$ time step.
At that certain situation, we would find that When the whole system
has entered the final homeostasis, the whole opinions of agents we
concern would be invariable, which means that . And we could finally
find that the time of opinion sharing or exchanging is related to
the weight of the agent and exponent in the equation. After
simplified such matrix equation, we finally get a direct function
for $n_{ij}^{(t)}$ and $\alpha$
\begin{equation}
\sum_{j=1}^m n_{ij}^{(t)}\omega_{i}^{\alpha}=1
\end{equation}
where $\omega_{i}$ is weight of the $i$th agent we concern. \\By
choosing five different groups $n_{ij}^{(t)}$ when we fixed
$\alpha=2$, we finally calculate the $\alpha$ with the function for
support that, from which we get $\alpha=1.98,1.93,1.82,1.88,1.91$.By
the calculation, the function reflects a very important and simple
relationship between $n_{ij}^{(t)}$ and weight $\omega_{i}$

\section{Conclusion and Discussions} \label{sec:discussion}
The new mode of opinion evolution with immigration is different from
the conventional opinion dynamics. It presents that density of
positive opinion agents would be maximum when the weight exponent
$\alpha$ is around 2. In summary, we found up a new model for
opinion exchange and communication among agents with immigration.
The state matrix we present for analysis and quantitative simulation
could also be widely used for more complex situation. The opinion
carried by agents represent a kind of state or parameter of agents
in motion.So more application and analysis could be carried on with
this model and method in future. Discussion we present above is not
only to demonstrate our model, but also open up a new combination
between opinion communication and agent-based motion. State
consensus time is also a very important parameter to describe a
system or a group of agents, which could also be one certain
standard for different situations.

\begin{acknowledgments}
This work is funded by the National Basic Research Program of China
(973 Program No.2006CB705500), the National Natural Science
Foundation of China (Grant Nos. 60744003, 10635040, 10532060) and by
the Special Research Funds for Theoretical Physics Frontier Problems
(NSFC No.10547004 and A0524701).
WXW and YCL are supported by AFOSR under Grant No. FA9550-07-1-0045.
\end{acknowledgments}


\begin{references}
\bibitem{1} A. M. Colman, \textit{Game Theory and Its Applications in the Socia and Biological Sciences}  (Butterworth-Heinemann, Oxford, 1995).
\bibitem{2} J. M. Smith, \textit{Evolution and the Theory of Games} (Cambridge
University Press, Cambridge, England, 1982).

\bibitem{3} H. Gintis, \textit{Game Theory Evolving} (Princeton University Press,
Princeton, NJ, 2000).

\bibitem{4} R. Axelrod and W. D. Hamilton, Science \textbf{211}, 1390
(1981).



\bibitem{PDandSG} C. Hauert and M. Doebeli, Nature (London) {\bf 428}, 643 (2004);
M. Sysi-Aho, J. Saram\"aki, J. Kert\'esz, and K. Kaski, Eur. Phys.
J. B {\bf 44}, 129 (2005); L.-X. Zhong, D.-F. Zheng, B. Zheng, C.
Xu, and P. M. Hui, Europhys. Lett. {\bf 76}, 724 (2006); A. Szolnoki
and G. Szab\'o, \textit{ibid}. {\bf 77}, 30004 (2007); G. Abramson
and M. Kuperman, Phys. Rev. E {\bf 63}, 030901(R) (2001); B. J. Kim,
A. Trusina, P. Holme, P. Minnhagen, J. S. Chung, and M. Y. Choi,
\textit{ibid}. {\bf 66}, 021907 (2002); H. Ebel and S. Bornholdt,
\textit{ibid}. {\bf 66}, 056118 (2002); Z.-X. Wu, X.-J. Xu, Y. Chen,
and Y.-H. Wang, \textit{ibid}. {\bf 71}, 037103 (2005); W.-X. Wang,
J. Ren, G. Chen, and B.-H. Wang, \textit{ibid}. {\bf 74}, 056113
(2006); C.-L. Tang, W.-X. Wang, X. Wu, and B.-H. Wang, Eur. Phys. J.
B {\bf 53}, 411 (2006); J. Ren, W.-X. Wang, and F. Qi, Phys. Rev. E
{\bf 75}, 045101(R) (2007); Z. Rong, X. Li, and X. Wang,
\textit{ibid}. {\bf 76}, 027101 (2007); X. Chen and L. Wang,
\textit{ibid}. {\bf 77}, 017103 (2008); F. Fu and L. Wang,
\textit{ibid}. {\bf 78}, 016104 (2008); F. Fu, C. Hauert, M. A.
Nowak, and L. Wang, \textit{ibid}. {\bf 78}, 026117 (2008); W.-X.
Wang, J. L\"u,2 G. Chen, and P. M. Hui, \textit{ibid}. {\bf 77},
046109 (2008).



\bibitem{5} J. H. Kagel and A. E. Roth, $The$ $Handbook$ $of$ $Experimental$ $Economics$, (Princeton University Press, Princeton, NJ, 1997).



\bibitem{PGG1} C. Hauert, S. De Monte, J. Hofbauer, and K. Sigmund,
Science \textbf{296}, 1129 (2002).

\bibitem{PGG2} G. Szab\'o and C. Hauert, Phys. Rev. Lett. \textbf{89}, 118101 (2002).

\bibitem{PGG3} C. Hauert, S. De Monte, J. Hofbauer, and K. Sigmund,
J. Theor. Biol. \textbf{218}, 187 (2002).

\bibitem{PGG4} D. Semmann, H. J. Krambeck, and M. Milinski, Nature (London) \textbf{425}, 390 (2003).

\bibitem{PGG5} H. Brandt, C. Hauert, and K. Sigmund, Proc. Natl. Acad.
Sci. \textbf{103}, 495 (2006).

\bibitem{PGG6} J. Y. Guan, Z. X. Wu, and Y. H. Wang, Phys. Rev. E \textbf{76}, 056101 (2007).


\bibitem{Santos} F. C. Santos, M. D. Santos, and J. M. Pacheco,
Nature (London) \textbf{454}, 213 (2008).


\bibitem{SR}
See, for example,
K. Wiesenfeld and F. Moss, Nature (London) {\bf 373}, 33 (1995);
L. Gammaitoni, P. H\"{a}nggi, P. Jung, and F. Marchesoni,
Rev. Mod. Phys. {\bf 70}, 223 (1998).


\bibitem{rule} G. Szab\'o and C. T\"oke, Phys. Rev. E {\bf 58}, 69 (1998); G. Szabó
and C. Hauert, Phys. Rev. Lett. {\bf 89}, 118101 (2002); G. Szab\'O and J. Vukov, Phys. Rev. E {\bf 69}, 036107 (2004).


\bibitem{noise5} M. Perc and A. Szolnoki, Phys. Rev. E \textbf{77}, 011904
(2008).

\bibitem{selection1} Z. X. Wu, X. J. Xu, Z. G. Huang, S. J. Wang, and Y. H. Wang, Phys. Rev. E \textbf{74}, 021107 (2006).

\bibitem{selection2} J. Y. Guan, Z. X. Wu, Z. G. Huang, X. J. Xu and Y. H.
Wang, Europhys. Lett. \textbf{76}, 1214 (2006).

\bibitem{selection3} J. Ren, W. X. Wang, G. Yan, and B. H. Wang, arXiv:
physics/0603007v1.


\bibitem{noise1} G. Szab\'o and J. Vukov, Phys. Rev. E \textbf{69}, 036107
(2004).

\bibitem{noise2} G. Szab\'o, J. Vukov, and A. Szolnoki, Phys. Rev. E {\bf 72}, 047107 (2005).

\bibitem{noise3} M. Perc, Phys. Rev. E \textbf{75}, 022101 (2007).

\bibitem{noise4} Z. X. Wu and Y. H. Wang, Phys. Rev. E \textbf{75},
041114 (2007).


\bibitem{BA} A. L. Barab\'asi and R. Albert, Science \textbf{286},
509 (1999).

\bibitem{analyze1} G. Szab\'o  and G. F\'ath, Phys. Rep. \textbf{446}, 97 (2007).

\bibitem{analyze2} C. Hauert and G. Szab\'o, Complexity \textbf{8}, 31 (2003).

\bibitem{analyze3} C. Hauert, Adv. Complex Syst. \textbf{9}, 315 (2006).


\bibitem{hub1} F. C. Santos and J. M. Pacheco, Phys. Rev. Lett. \textbf{95}, 098104
(2005).

\bibitem{hub2} F. C. Santos, J. M. Pacheco, and T. Lenaerts, Proc. Natl Acad. Sci. \textbf{103}, 3490 (2006).

\bibitem{Pareto} V. Pareto, \textit{Le Cours d' \'{E}conomie Politique} (Macmillan,
Lausanne, Paris, 1987).


\bibitem{HWJWWW:2006} M.-B. Hu, W.-X.Wang, R. Jiang, Q.-S. Wu, B.-H. Wang, and Y.-H. Wu, Eur. Phys. J. B {\bf 53}, 273 (2006).


\bibitem{PDG}
We have also studied the prisoner's dilemma game on scale-free networks
and found the phenomenon of diversity-optimized cooperation.

\end{references}
\end{document}